\documentclass[fleqn,usenatbib]{mnras}

\usepackage{newtxtext,newtxmath}

\usepackage[T1]{fontenc}
\usepackage{ae,aecompl}

\usepackage{graphicx}	
\usepackage{amsmath}	
\usepackage{color} 
\usepackage{multirow} 
\usepackage{hyperref}

\title[New approaches for faint source detection in hard X-ray surveys]{New approaches for faint source detection in hard X-ray surveys}

\author[V. A. Lepingwell et al.]{
V. A. Lepingwell,$^{1,2}$\thanks{E-mail:V.A.Lepingwell@soton.ac.uk}
A. J. Bird,$^{1}$
and
S. R. Gunn,$^{2}$
\\
$^{1}$School of Physics and Astronomy, University of Southampton, University Road, Southampton, SO17 1BJ, UK\\
$^{2}$School of Electronics and Computer Science, University of Southampton, University Road, Southampton, SO17 1BJ, UK}

\date{Accepted XXX. Received YYY; in original form ZZZ}

\pubyear{2018}

\begin{document}
\label{firstpage}
\pagerange{\pageref{firstpage}--\pageref{lastpage}}
\maketitle

\begin{abstract}
We demonstrate two new approaches that have been developed to aid the production of future hard X-ray catalogs, and specifically to reduce the reliance on human intervention during the detection of faint excesses in maps that also contain systematic noise. A convolutional neural network has been trained on data from the INTEGRAL/ISGRI telescope to create a source detection tool that is more sensitive than previous methods, whilst taking less time to apply to the data and reducing the human subjectivity involved in the process. This new tool also enables searches on smaller observation timescales than was previously possible. We show that a method based on Bayesian reasoning is better able to combine the detections from multiple observations than previous methods. When applied to data from the first 1000 INTEGRAL revolutions these improved techniques detect 25 sources (about 5\% of the total sources) which were previously undetected in the stacked images used to derive the published catalog made using the same dataset. 

\end{abstract}




\section{Introduction}



Thanks to the current generation of space telescopes with survey capabilities, the sky in the hard X-ray / soft gamma-ray band (approximately 10keV-1MeV energy) has shown itself to be both well-populated and highly variable. Surveys in this band are normally carried out using coded aperture telescopes that provide a good sensitivity across a wide field of view (typically $>100$ square degrees), and as such allow frequent returns to the same sky region, producing rich datasets with information in both spatial and temporal dimensions. 

However, the analysis of data from coded aperture telescopes is not trivial, as it is an indirect imaging method and sky images can contain systematic noise as a result of an imperfect instrument model, and also when an adequate description of the source distribution cannot be determined, as is sometimes the case in crowded regions where sources cannot be fully resolved. The sources in the hard X-ray sky display a huge dynamic range, and are thus detectable on many different timescales. While the brightest sources can be detected in a single observation, the faintest sources may require 1000s of images to be co-added. More recent surveys \citep{Bird2010, Bird2016} have searched for ways to detect sources on all timescales in an efficient way, but these are generally expensive in operator effort, and as the data from such surveys is ever increasing, there is a need for automated techniques that can scale with the data when it exceeds the capacity to be processed manually.

\begin{figure*}
\includegraphics[width=1.0\textwidth]{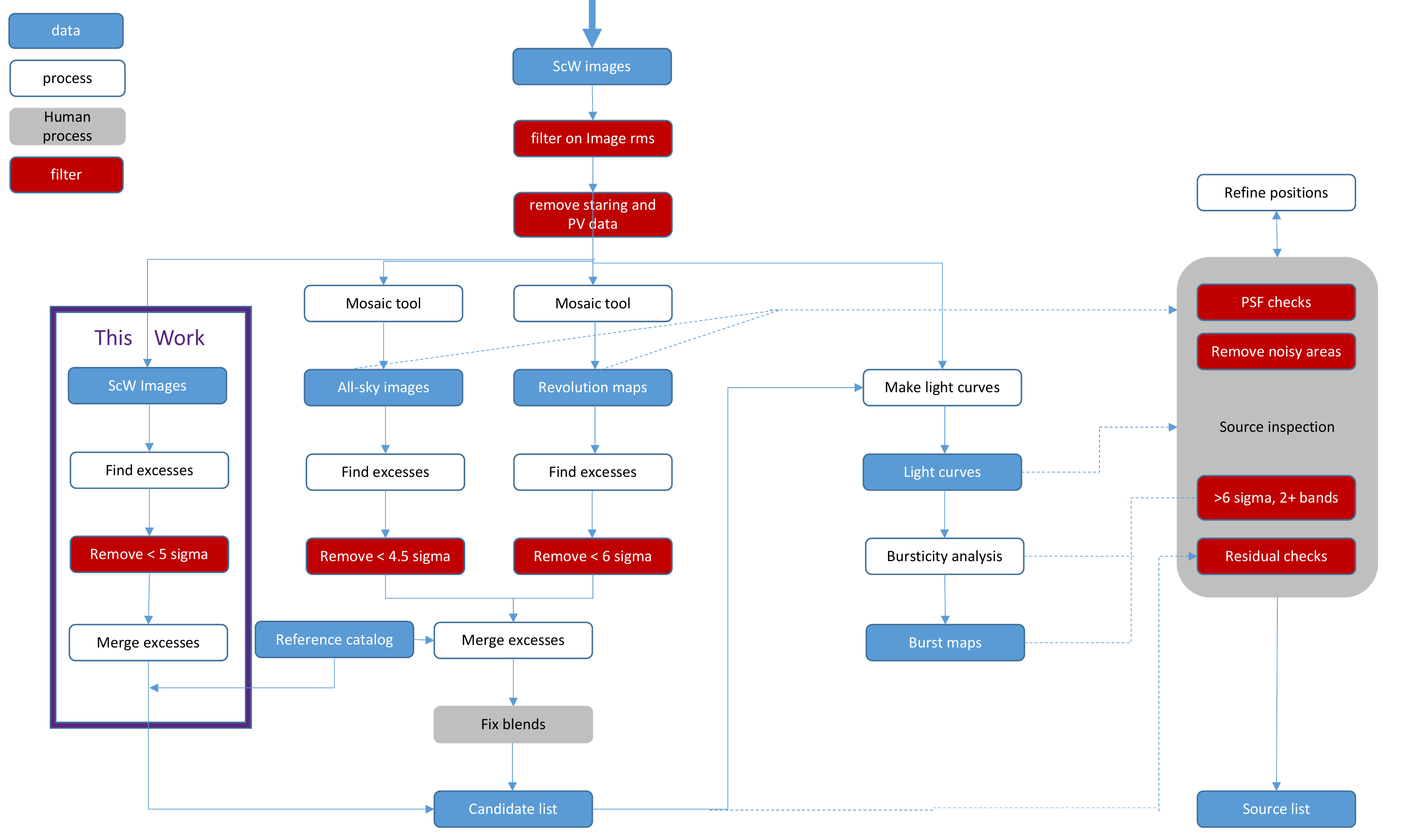}
   \caption{Data analysis and source selection flowchart, showing the filtering criteria applied at each stage during the cat1000 process \citep[adapted from][]{Bird2010}. The steps that are distinct to this work are shown in the highlighted box to left. While the general chain of steps is similar, this work is distinct in several ways: (1) source detection is done at a ScW level and uses a CNN trained to classify 11x11 pixel windows rather than search for sources across an entire image, (2) excesses are merged to each other in a Bayesian fashion that prioritises by goodness of cluster rather than chronological discovery date, and (3) only merges in a reference catalog as the final stage in the process to avoid introducing initial biases.}
   \label{fig:flow}
\end{figure*}

In this paper, we look at new methods to efficiently search for excesses in sky maps that may also contain systematic artefacts of the imaging process, and how to intelligently combine lists of excesses found in multiple maps. These maps may be generated in different energy bands, in different sky orientations, or at different times - but all may be considered potential information concerning each putative source. Due to the dynamic range across the source population, a bright persistent source may be detected 1000s of times in individual observations (as well as in any co-added maps) whereas information on a fainter source may be determined from just co-added images. Conversely a bright transient may be seen in just a few observations, and be completely undetectable in a co-added image mosaic. The challenge is to combine all the excess detections into a coherent source catalog, suppressing statistical noise in the presence of systematics effects, efficiently and without introducing the subjective biases that human intervention can produce. 

Specifically for this study, we use the images produced by the INTEGRAL/IBIS telescope in the $\sim$18-100 keV band, but the methods would be equally applicable to the Swift/BAT telescope images.

The European Space Agency's (ESA) International Gamma Ray Astrophysics Laboratory (INTEGRAL) was launched in 2002 and has performed over 18 years of observations in the energy range 5 keV to 10 MeV.  It has a complement of three primary high energy instruments: the SPectrometer on INTEGRAL (SPI), the Joint European X-ray Monitor (JEM-X), and the Imager on Board INTEGRAL (IBIS). IBIS is composed of the INTEGRAL Soft Gamma Ray Imager (ISGRI) and Pixellated Imaging CaeSium Iodide Telescope (PICsIT) detectors surrounded by an active veto. IBIS/ISGRI uses a tungsten coded mask to determine source astrometry through de-convolution of the mask pattern (shadow) projected onto the detectors.

IBIS (Imager on Board the INTEGRAL spacecraft) \citep{Lebrun2003} in particular has been optimised to produce data for surveys \citep{Ubertini2003} as it has a large (30 degrees) FOV (field of view) and good resoultion thanks to the coded mask. Survey catalogs exploiting data obtained from ISGRI have been published at regular intervals \citep{Bird2004, Bird2006, Bird2007, Bird2010}. The most recent, `Catalog of 1000 orbits' \citep[hereafter cat1000][]{Bird2016}, is an all-sky, soft gamma-ray source catalog which uses data from INTEGRAL's first 1000 orbits. This survey had a total exposure time of approximately 3.5 years and included 939 sources. This catalog used a light-curve based method to search for transient events, but did not attempt to search for sources on the shortest timescales due to the size of the dataset that would have created. The shortest timescale mapped corresponded to the $\sim$3 day orbit of INTEGRAL.
Other INTEGRAL-based catalogs have been produced, and focus on specific sky areas such as the Galactic Plane \citep{Krivonos2010, Krivonos2012} and generally only use co-added maps of the dataset in search of the faintest persistent sources. JEM-X and SPI also use coded masks, and observations from these were used to produce a catalog in 2007 and 2008 \citep{westergaard2007, Bouchet2008}.

The techniques employed to generate ISGRI catalogs (such as cat1000) are no longer adequate as they do not scale well to the ever increasing data set. Cat1000 took 9 domain experts to spend 2.5 years to produce and as INTEGRAL is now approaching 2500 orbits there is a clear need to improve the techniques and tools used.

This paper introduces two new tools we have developed using deep learning and Bayesian reasoning to improve how we search multiple ISGRI maps for sources and combine these detections to define catalog sources, respectively. We will first introduce the data set, then a description of the architecture of the deep learning method and the training process including how the test and training set were generated is presented. We compare this new method to the source detection tools used in producing recent catalogs.  We then present an improved merging algorithm based on \citet{Budav2008} used to combine excess detections from multiple maps and compare this new method to the one previously used in generating ISGRI catalogs.   Figure \ref{fig:flow} shows how the previous catalogs have been produced and how these new tools would complement the process.  These methods not only reduce the human time needed to create these catalogs but also allow us to search the ISGRI maps on a science-window (ScW) scale, which until now was not feasible.

Cat1000 used both the standard astronomy package SExtractor \citep{Bertin1996} and a custom-made piece of software called {\em peakfind} to detect sources in stacked ISGRI maps. We have developed a new tool using a deep learning technique -- specifically, a convolutional neural network (CNN) -- to search ISGRI maps to detect sources. This paper will compare our new tool to the traditional methods for source detection. 

Cat1000 used a custom-built piece of software called {\em megamerge} to combine the detections found in multiple maps. We have employed a new method using Bayesian matching \citep{Budav2008} that removes some of the bias that was inherently part of the {\em megamerge} process. 

This paper is organised as follows. Section \ref{sec:data} provides a brief summary of the ISGRI data, and Section \ref{sec:old_detection} summarises the previous methods for detecting sources. We briefly introduce CNNs and describe our new CNN-based approach to source detection in Section \ref{sec:cnn}. We then discuss previous methods for merging detected sources in Section \ref{sec:old_merging} before describing our new Bayesian matching technique in Section \ref{sec:new_merging}.  We analyse the performance of our tools in Section \ref{sec:performance} and summarise our results in Section \ref{sec:conclusions}.

\section{The Dataset}
\label{sec:data}

In order to develop these two new tools and also to test their performance against the traditional catalog tools, we have used the original cat1000 data set, which is all publicly available INTEGRAL/ISGRI data obtained up to the end of 2010. INTEGRAL has an orbit (or revolution) of approximately three days and each revolution is broken down into Science Windows (ScWs), short pointings which have a duration of $\sim$ 2ks. ScW images have a pixel size of 4.8 arcminutes (0.08 degrees), and the typical resolution of the instrument is 12 arcminutes.  As in cat1000, we include all revolutions from 45 onwards and all pointed observations up to revolution 1000 (December 2010) unless flagged as Bad Time Intervals (flagging is provided by the INTEGRAL Science Data Centre, ISDC) for a total of $\sim$ 67000 ScW. Revolutions up to 45 were excluded from cat1000 as data from the instrument up to this point were extremely noisy. 

\section{Previous method for ISGRI Source Detection}
\label{sec:old_detection}
In cat1000 approximately 67000 ScWs were used to create stacked maps  using a purpose-built image mosaic tool which was developed to statistically average the images from multiple input maps. This allowed all-sky maps (these all-sky maps have a pixel size of 2.4 arcminutes) to be created from a large number of input ScWs. Figure \ref{fig:flow} shows the entire cat1000 process from the ScW images to the final sources list. Mosaics were constructed for five energy bands and in four sky projections: centred on the Galactic Center, on the Galactic anti-center, north Galactic polar and south Galactic polar. These projections were chosen in order to reduce PSF distortion which impedes the source detection algorithms used.

60 all-sky maps and over 19000 revolution maps were constructed and searched for flux excesses indicating an astrophysical source to produce an initial excess list. Two different techniques were used to search the mosaics maps, the standard SExtractor tool, and {\em peakfind} which was developed specifically for use on ISGRI maps which takes into account the varying levels of systematic background. SExtractor has sophisticated algorithms for pre-filtering the image to enhance detection of specific PSFs, and is capable of de-blending some complex source regions, both of which are important for ISGRI maps. However, it does not work so well with local variations in systematic noise levels, which are a feature of the ISGRI maps. {\em peakfind} uses a recursive search around the peak position to detect excesses, maintaining some de-blending capability but performing only limited tests on the shape of the PSF detected. The main benefit of {\em peakfind} is that it performs a local assessment of the image rms and assesses excesses relative to that local background. As such it is much less vulnerable to over-detection in noisy areas of the maps.

The use of multiple excess detection algorithms was valuable in cat1000 production as the different methods used could be compared and an excess appearing in both lists could be treated with higher confidence. However the different underlying approaches meant that all excesses still needed manual checking as complex regions were often interpreted differently by the two algorithms, and this was a natural path for operator bias to be introduced. 

During cat1000 production, both SExtractor and {\em peakfind} had to be run on a large number of maps in five individual energy bands, for different sky projections and on different timescales - revolution level, observation sequences and whole-archive. Although the performance of the methods was similar, the time taken for this and the subsequent combination of the excesses found in these maps was a complex task. 

For future catalogs, it is hoped that new techniques can be developed to make this task more tractable - and indeed extend it to the ScW timescale data which has not been attempted so far. The use of HEALPix (Hierarchical Equal Area isoLatitude Pixelation) based maps will reduce the number of sky projections needed, and an image search that combines multiple energy bands will not only save time but may also provide a more robust detection as the energy bands are not completely independent and we would expect a source appearing in one energy band to appear in some, but not necessarily all, others. Such combination logic was part of the manual inspection of the excesses, and an automated method which took the same approach should be less vulnerable to random image noise. Unfortunately this is still not fool-proof, as systematic noise and ghost sources appear at the same position in every energy band. In principle, an automated method could also recognise sensible ratios of flux in different energy bands as the typical X-ray source spectra give rise to fairly predictable fluxes across the energy bands.

\section{Deep Learning Method}
\label{sec:cnn}
\subsection{Convolutional Neural Networks}
Convolutional neural networks (CNN) developed in recent years are most commonly applied in the field of image processing because they perform well at dealing with image recognition and classifications tasks and are considered to be one of the leading techniques in the field. \citep{Lecun1995}. In the absence of domain knowledge they can work well with raw features; a CNN automatically learns the underlying features required to detect when a source is present \citep{Schmidhuber2015}. These distinct advantages make a CNN the ideal choice for source detection in high energy astronomy images. A CNN is a supervised method and thus requires a training set to be run through a CNN many times, adjusting the CNN's parameters using backpropagation to minimise a loss function \citep{lecun1988}.

In image classification, pixels that are near each other are quite likely to be more related than two pixels that are further away. This means that the pixels' proximity to one another is an important factor whilst classifying and CNNs specifically take advantage of this fact \citep{Lecun2015}. 
In a standard neural network, every pixel is linked to every single neuron, in the case of image classification this added computational load makes training more difficult and resulting models are often less accurate. A CNN removes a lot of these less significant connections, and makes the image processing computationally manageable through filtering the connections by proximity. In a given layer, rather than linking every input to every neuron, CNNs restrict the connections intentionally so that any one neuron accepts the inputs only from a small subsection of the previous layer. Therefore, each neuron is responsible for processing only a certain portion of an image. Combined with effective training methods for deep layer networks this provides a powerful approach.
An image in a CNN is passed through a combination of successive layers, every channel of the image is presented to the network at once:
\begin{itemize}
\item {\textit{Convolutional layer}: here the filters can be thought of as feature identifiers.}
\item {\textit{Nonlinear layer}: the CNN uses ReLu (negative input are zeroed) \citep{Hahnloser2000}. This allows the network to approximate arbitrary functions by introducing non-linearities. \citep{Nair2010}}
\item{\textit{Max Pooling layer}: this down-samples the input, thereby increasing the network's efficiency and allowing the network to train quicker \citep{Nagi2011}} 
\item{\textit{Fully connected layers}: neurons in a fully connected layer have connections to all activations in the previous layer. The CNN can then use a softmax activation function to produce the final output: a probability of the input being a source and not a source (and ensure a partition of unity) \citep{Bishop2006}.}
\item{\textit{Categorical crossentropy
}: This is a loss function that is used in multi-class classification tasks where an example can only belong to two or more label classes, and the model must decide which one. It computes the loss between the labels and predictions. \citep{zhang2018}.}
\end{itemize}


A CNN is a very powerful and efficient model which, unlike some other machine learning methods, performs automatic feature extraction. The network picks out the important features in an image in order for it to make highly accurate classifications. In fact CNNs can outperform humans in image classification due to the networks' ability to pick out underlying patterns and structures that domain experts can be unaware existed.

\subsection{Training the CNN}
\begin{figure*}
\includegraphics[width=1.0\textwidth]{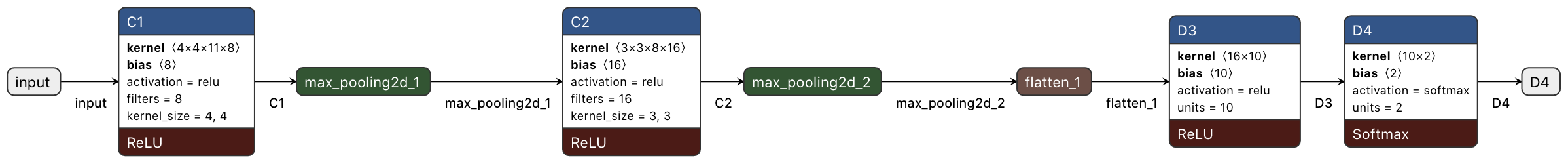}
   \caption{The architecture of the final source detector CNN. The CNN uses two convolutional layers with kernels of various sizes, followed by a multi-layer perceptron and softmax function to determine final classification of the image as containing a source or not. A visual representation of the CNN applied to an example image is presented in Figure~\ref{fig:CNN_Architecture_filters}.}
   \label{fig:CNN_Architecture}
\end{figure*}
Fundamentally, our source detection approach is an image classification task where the CNN learns the features which separate a window containing a source from the background. Unlike other source detection approaches, the CNN does not require the entire image because the 11x11 pixel windows contain sufficient information to train a CNN to distinguish source from background. To enable the CNN to have enough information to classify we needed to train it with labelled examples of both sources and background.

The CNN was trained using $\sim$25,000 examples of both sources and background extracted in $11\times 11$ pixel windows from 11 channels: intensity and significance across five energy bands (17-30, 30-60, 18-60, 20-40 and 20-100 keV), plus the exposure map. All the images from the five energy bands are stacked and sent through the network at once. This allows the CNN to be able to see a whole spectrum of information about the source at once. we also have included the exposure map as our 11th channel. This gives the CNN information about how near to the centre of the map a window is as in an ISGRI ScW image, as the centre of the image has the highest exposure and the edges of the map the least. 

Once the network was fully trained, a $11\times 11$ pixel window could be moved across an entire ISGRI ScW image which could contain many sources. Figure \ref{fig:CNN_Architecture_filters} shows how the network takes the input window image, passes it through a series of convolution, nonlinear activation (ReLu), pooling (downsampling), and fully-connected layers and then returns an output for each window, with the assumption that each window contains only one source. In all but the galactic centre, sources are sparse enough that you would have a very low chance of detecting more than one in a $11\times 11$ pixel window.

To generate our training set we used $\sim$200 ScWs ($\sim$a fifth of the ScWs in cat1000) and in each ScWs used the results file generated from the IBIS pipeline \citep{Goldwurm2003} to select any sources that were present with a significance of over 5 sigma. Any machine learning model trained on a human-labelled dataset could potentially learn the biases of the labeller. The IBIS pipeline uses The General Reference Catalog as the master table listing all known high-energy sources of relevance to INTEGRAL. These objects are all those that have been detected by INTEGRAL or that are known to be brighter than 1mCrab in the 1keV to 10MeV band. The objects in the catalog were compiled from several sources so we have confidence in the validity of these training examples and that any individuals' bias would have limited impact.

For every source we also selected a random window of background to include in the training set. The resulting training set contained $\sim$ 25,000 examples of source and background. 70 percent of this was used to train the CNN while 30 percent was left out of the training processes and used as a test set to measure performance. 

\subsection{CNN architecture }
\begin{figure*}
CNN architecture
\includegraphics[width=0.8\textwidth]{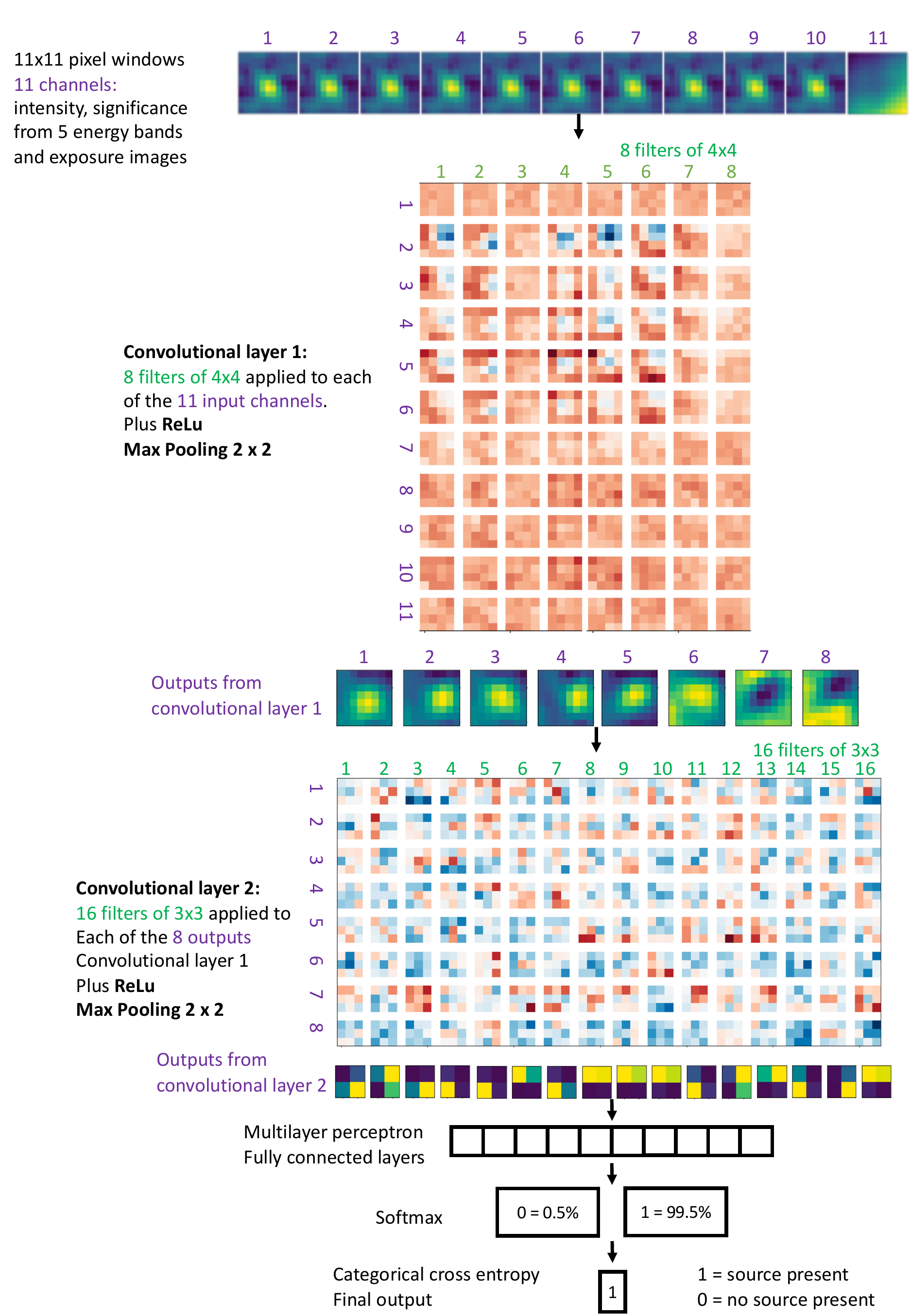}
   \caption{The architecture of the source detector CNN, which shows how a $11\times 11$ pixel window with a source present is processed through the CNN resulting in its detection. The first convolutional layer applies eight 4x4 filters to the pixels windows from each of the 11 energy bands producing the feature map shown. A max pooling layer and ReLu function applied to these maps produce the eight outputs from the first convolutional layer shown as the centre row of images. The next convolutional layer applies a similar sequence of processing, as do any subsequent layers in a CNN. The outputs from the convolutional layers are passed through a multilayer perceptron and a softmax function to decide the final classification of the image as containing a source or not.}
   \label{fig:CNN_Architecture_filters}
\end{figure*}
We trained several CNN networks with different architectures with the aim of designing the simplest architecture as possible without effecting the network's performance. We found that by increasing the number of filters and layers from the architecture chosen and shown in figure \ref{fig:CNN_Architecture} there was no improvement in performance but when we reduced the number of filters we saw a noticeable change in performance. For example a network with six filters in the first convolutional layers and 14 in the second convolutional layer found 6 false negatives and 4 false positives when applied across the test set, this is in contrast to just 2 false negative and no false positives from the final network.  Figure \ref{fig:CNN_Architecture_filters} breaks down our network layer by layer and illustrates how an 11x11 pixel window that includes a source would be passed through the network. 

\subsection{Testing the CNN}

\begin{figure}
\includegraphics[width=\linewidth]{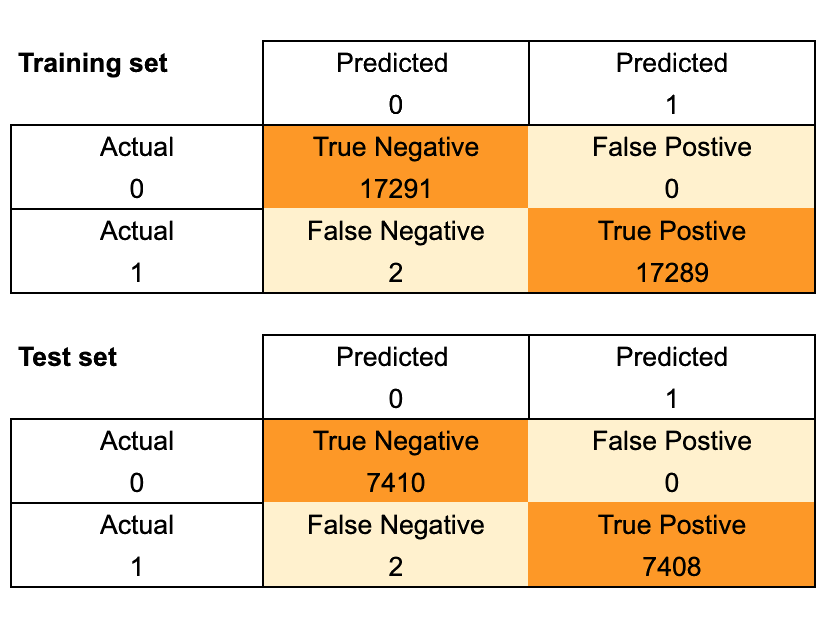}
   \caption{Confusion matrices for the data sets used to train and test the final CNN. For both data sets, each row represents the truth for the labelled 11x11 pixel windows, while each column corresponds to the predictions of our CNN. Correctly classified sources (shown in dark orange boxes on the diagonal) dominate over misclassifications (light yellow boxes on off-diagonal).}
   \label{tab:con_mat}
\end{figure}

Figure \ref{tab:con_mat} show the confusion matrices for the CNN for both the training and the test set. From all these 11x11 pixel windows none of the background sources were classified incorrectly and only 4 sources were not detected. 

We also trained another five CNNs with the same architecture, with similar performances, each using a different fifth of the ScWs used in cat1000. This was to ensure we had produced a stable network that would have a similar high level of performance regardless of which ScW were used in the training stage and that the model was not overfitting.

The CNN takes $\sim$ 6 hours to train using Keras with a TensorFlow backend on a NVIDIA TITAN Xp with 12 gigabits of ram and took $\sim$1 day to apply to the whole cat1000 dataset. 

\section{Previous method for Merging Excesses }
\label{sec:old_merging}
SExtractor and {\em peakfind} will detect sources in every map, so every source will most probably be detected multiple times across the entire dataset. A method to merge the excesses into a unique set of sources is therefore required. An algorithm called {\em megamerge} iteratively merges the excess lists from each map into a base list which takes the cleaning catalog (only the sources in the ISGRI reference catalog that have been previously detected by IBIS) as a starting point. It should be noted that SExtractor and {\em peakfind} are applied to each of the five energy bands individually. 

The first drawback from using {\em megamerge} is that as it uses the reference catalog as a starting point for the merging it adds a bias into the process. This can be compounded when the source was originally discovered by INTEGRAL. A further problem is that the result is dependent upon the order in which the excesses to merge are presented to the algorithm. New excesses are presented to the merging algorithm and a decision on whether to merge with an existing excess in the database is made purely on position - on whether the two excesses lie within a given merge radius. The `catalog' position of the excess may be updated if the newly merged excess has higher significance than those previously used - the point source location accuracy of a coded mask imager depends strongly on the detection significance. The inherent risk though is that a strong new detection at the limit of the merger radius can not only `steal' an existing detection but also cause its coordinates to change. Clearly the order in which the excesses are presented can have a significant influence on the outcome of the process. No attempt was made to optimise or mitigate this during cat1000 production, so as a result significant effort was required to check results manually.

The {\em megamerge} algorithm makes no use of pre-existing information (beyond the position) to decide if two sources separated by the merger limit are likely matches. As a simple example, two 50-sigma excesses separated by 8' are very unlikely to be the same source, but would be erroneously merged by {\em megamerge}. On the other hand, two 4.5-sigma excesses separated by 8' may be the same source, since the point source location accuracy at $50\sigma$ and $5\sigma$ are $\sim$0.5' and $\sim$5', respectively \citep{Scaringi2010}.

As a result of these limitations, and concerns over how robust the outcome was, we searched for a better approach to merging the source lists.

\section{Bayesian Matching}
\label{sec:new_merging}
We have tailored a method used by XMM to work on ISGRI data \citep{Rosen2016}. This method first merges individual observations into points using Bayesian probability before matching the merged point to a source in the reference catalog - this removes the bias we noted in {\em megamerge} as the reference catalog is no longer being used as a starting point and the order the excesses are presented to the algorithm are now irrelevant as it uses Bayesian probability to decide which order to merge in. 

First the algorithm searches for any pairs of ScW detections that are less than 8 arcminutes away. Any detections found within the same ScW are excluded from this matching process as they will be different sources. Each pair has a Bayesian match probability $(p_{match})$ assigned to it using equation 1 where $\sigma_1$ and $\sigma_2$  are the position error radii of each detection in the pair (radians), $\psi$ is the angular separation between the pair, $p_o$ = $N^*/N_1N_2$ where $N_1$ and $N_2$ are the number of objects in the sky based on the surface densities in the two fields. Each $N$ value is derived from the number of detections in the two observations and then scaled to the whole sky and $N^*$ = number of objects common between them.

This allows the algorithm to make a first cut and remove any pairs with $p_{match} < 0.5$ then determine the order to match the detections. A goodness of cluster (GoC) is also used to do this by prioritising the detections that have the smallest error radius and most amount of pair matches. The algorithm sorts the excesses in order of ascending GoC and iterates down the list. For each excess the algorithm sorts that excess's matching pairs in ascending order. We then progress down this GoC-sorted list to merge sources by assigning a common source ID to sources matched in a pair. The process concludes when each cluster has an unique source ID. 

\begin{eqnarray}
    p_{match} = \left[1 + \frac{1-p_0}{B\cdot p_0}\right]^{-1}
	\label{eq:bayesian_matching} \\
	B = \frac{2}{\sigma_1^2+\sigma_2^2}\exp{- \left[\frac{\psi^2}{2(\sigma_1^2+\sigma_2^2)}\right]}
\end{eqnarray}



\begin{table*}
\begin{tabular}{ccccc}
\hline
\hline
Detection method            & Excesses found over 5 sigma                & Merging method    & cat1000 sources & non cat1000 sources \\ 
\hline
\hline
\multirow{2}{*}{CNN}        & \multirow{2}{*}{$\sim$100,000} & Megamerge         & 434            & 14                 \\
                            &                                & Bayesian Matching & 448            & 25                 \\
\hline
\multirow{2}{*}{SExtractor} & \multirow{2}{*}{$\sim$200,000} & Megamerge         & 93             & 1                  \\
                            &                                & Bayesian Matching & 96             & 2                  \\
\hline
\multirow{2}{*}{Peakfind}   & \multirow{2}{*}{$\sim$500,000} & Megamerge         & 175            & 0                  \\
                            &                                & Bayesian Matching & 179            & 5 \\
\hline
\end{tabular}
\caption{Sources detected by each method. CNN detections utilise all five energy bands simultaneously and produce a single detection for each source, whilst SExtractor and {\em peakfind} search each energy band individually meaning a single source could be detected separately in all five energy bands (which ultimately requires significantly more time to perform the merging process).
}
\label{tab:performance}
\end{table*}

\section{PERFORMANCE}
\label{sec:performance}
To measure the performance of our new tools we have applied them to the cat1000 dataset, and we have also applied the traditional tools. In this section we discuss the outputs from each method and how they compare. We also provided a list of detections to domain experts for manual inspection. Also we take a detailed look at the galactic centre as the most difficult region to catalog in cat1000. 

It is important to note that comparisons between this work and cat1000, while instructive, are not meant to be direct comparisons of exactly parallel approaches. Indeed, our ScW-based search is ideal for finding significant transient events on small timescales and strong persistent sources, while weaker persistent sources are better suited for discovery in stacked images such as those analysed for cat1000.

\begin{figure*}
\includegraphics[width=0.95\textwidth]{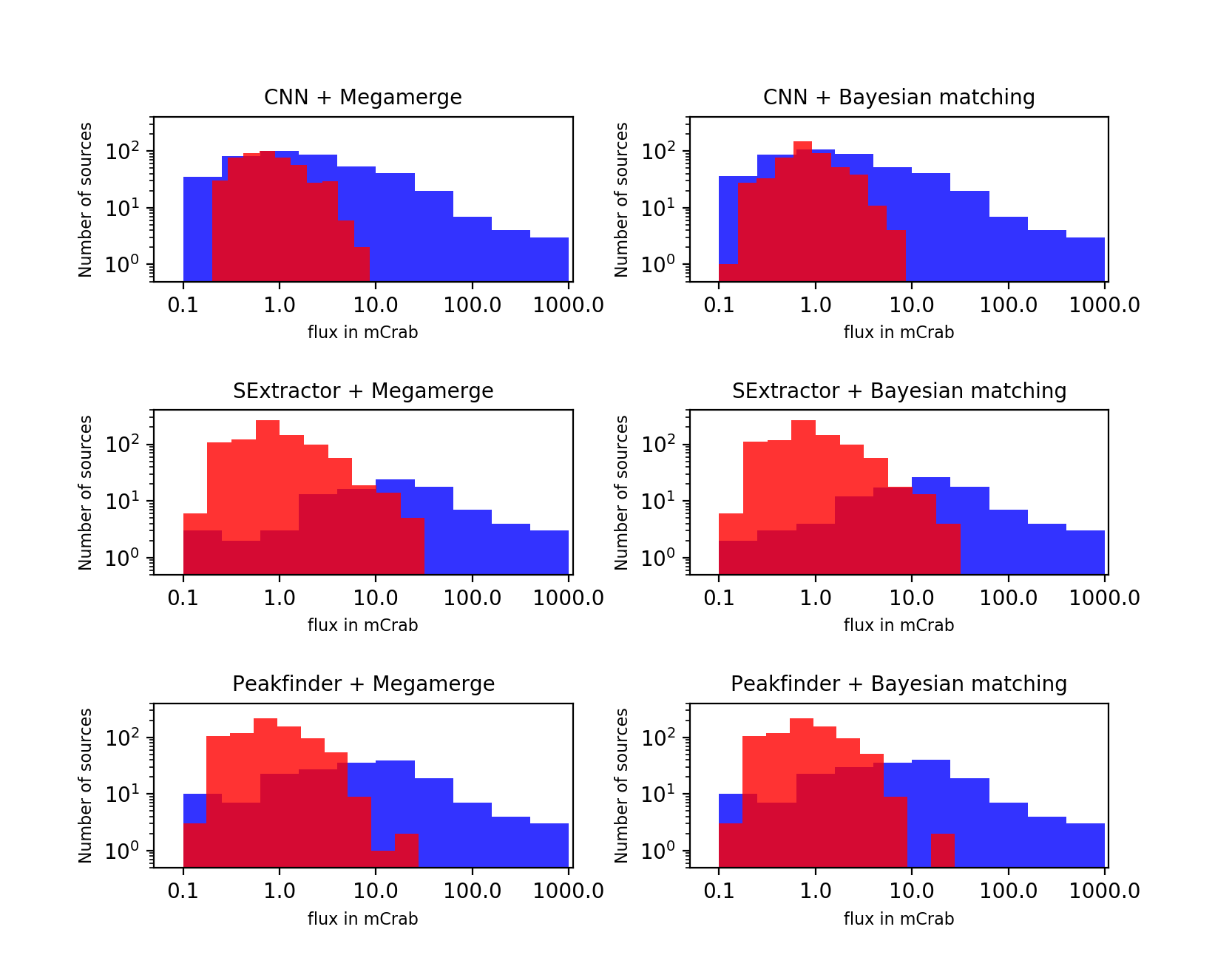}
   \caption{Histograms showing the proportion of cat1000 sources found (blue) vs not found (red) and their fluxes in cat1000 for every combination of the three source detection techniques and the two merging techniques. While other source detection approaches produce a detection efficiency that decreases (generally) monotonically as source brightness decreases, our method is able to detect fainter sources by leveraging the subtler source signatures captured by the CNN. These sources can have exceptionally low average fluxes (0.1 mCrab) in the stacked images used for cat1000, but will be bright enough in some ScW images to be detected using these methods.}
   \label{fig:detection histo}
\end{figure*}

\subsection{Method Comparison}

Each of the three source detection methods was applied across the entire data set to generate an excess list. While {\em peakfind} and SExtractor were applied to each energy band separately, our CNN utilised the five bands simultaneously. Thus the CNN finds about 1/5 as many excesses as {\em peakfind}, which makes similar judgements about the PSF as the CNN (SExtractor does not perform as well in this regard) These three excess lists were then passed through both merging methods to generate a list of merged sources. Table \ref{tab:performance} breaks down each method and shows how many cat1000 sources were found and also how many sources were found that were not included in cat1000. Not only do the new CNN and Bayesian matching methods recover more sources, but figure \ref{fig:detection histo} shows how these methods have a lower flux threshold compared to the traditional tools. Overall the CNN can detect sources at a lower flux than the other two methods.

All of the methods presented here find fewer sources than the full cat1000 due to the fact we are applying these methods on single ScW images, whereas cat1000 was applied to stacked images. These stacked images were ideal for finding faint persistent sources, which is not possible on a ScW level. ScW-level analysis finds bright persistent sources or fainter transient sources, and our CNN performed best at finding such sources missed in cat1000.

\begin{figure}
\includegraphics[width=\linewidth]{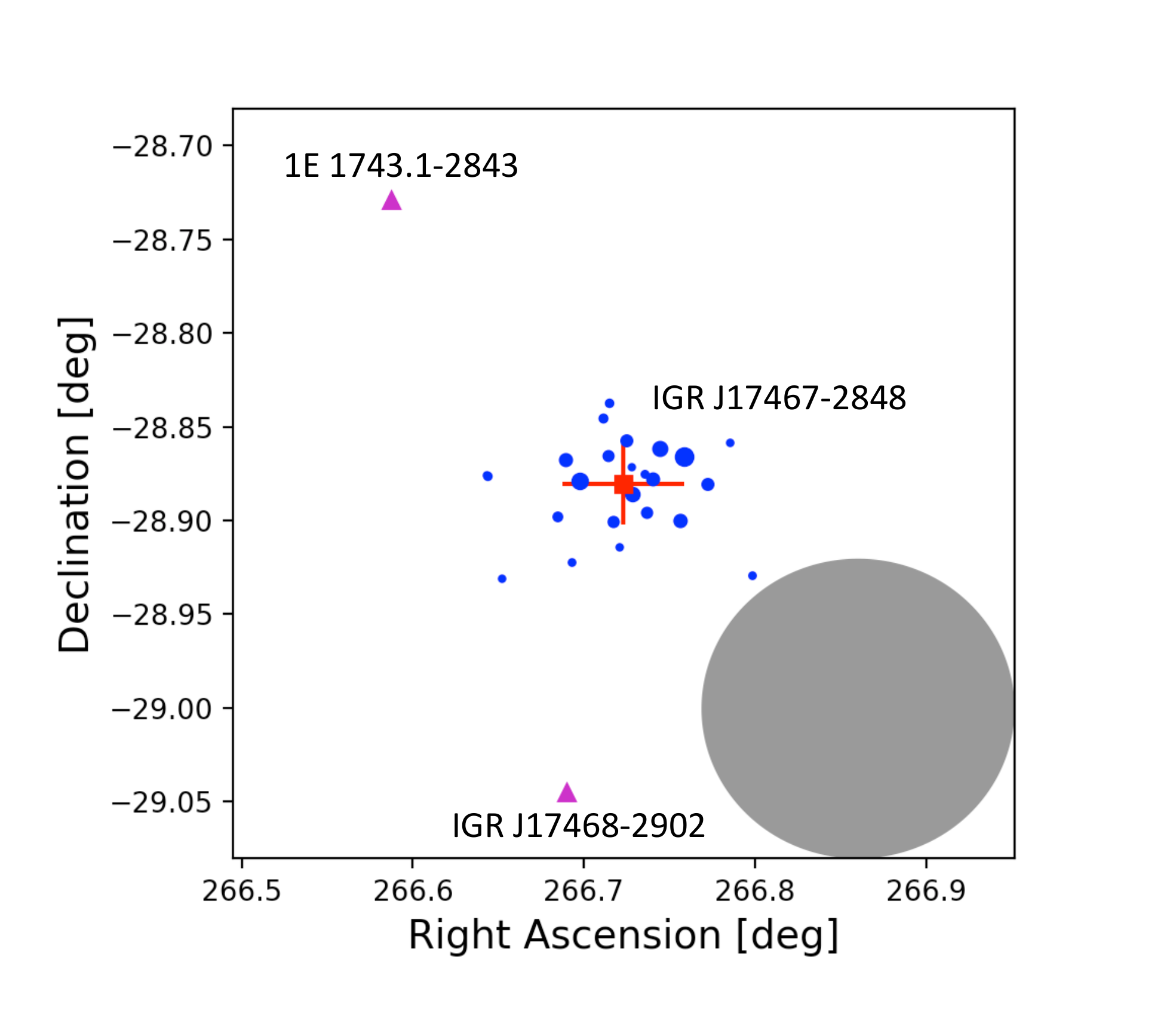}
\caption{Our detections of a source (IGR J17467-2848) detected with our method but missed in cat1000. Blue circles denote our detections, with marker size indicating detection significance, and the weighted final position of the source (with its uncertainty) is shown as a red square with error bars. Other nearby cat1000 sources are shown as magenta triangles, and the resolution of ISGRI is shown as the large grey circle.}
\label{fig:IGR}
\end{figure}

It is noteworthy that 25 sources that were not included in cat1000 were found using the new CNN and Bayesian matching methods. An example of one of these sources, IGR J17467-2848
is shown in figure \ref{fig:IGR}. These sources were presented to domain experts that were part of the cat1000 survey team for them to manually inspect the sources and provide an evaluation of the performance of the CNN. Out of the 25 sources the domain experts agreed that all but two were sources. These two sources were both found in the noisy borders of the ISGRI images which exist due to the coded-mask of the instrument. A member of the survey team would reject this on manual inspection but due to the localised nature of the CNN it would have been hard for the CNN to have picked these out as noise. To avoid this problem in the future, the CNN could either be just applied to a small section of the ScW image that avoided this area, or these detections could be flagged for manual inspection. While the addition of the exposure map as the 11th channel significantly reduces this effect, it does not entirely remove all artefacts from the extreme edges of the image.

\subsection{Blended Sources and Galactic Centre}
\begin{figure*}
\includegraphics[width=0.9\textwidth]{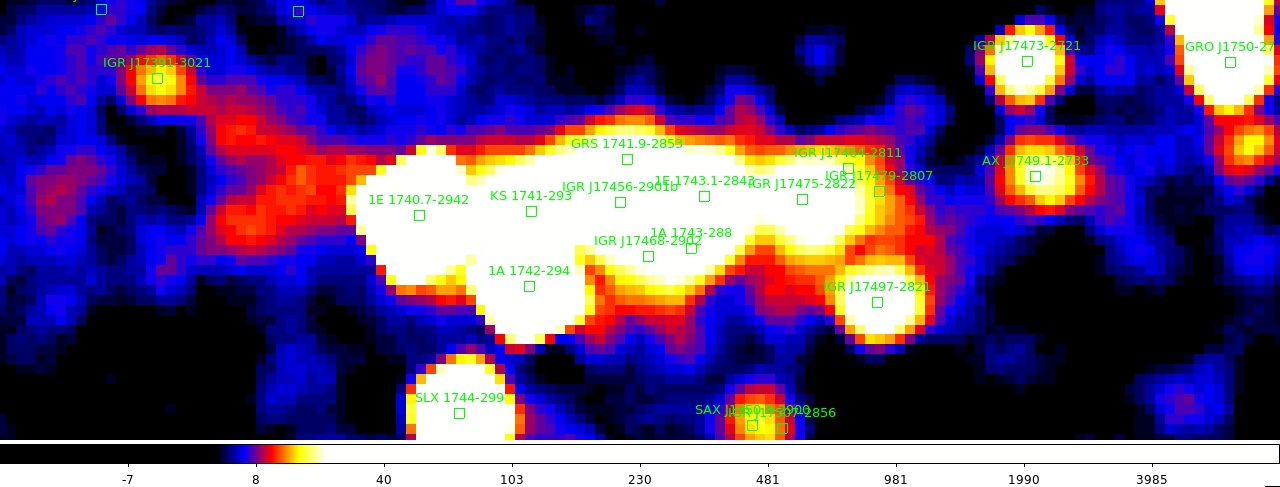}
\caption{\{Galactic centre region of stacked ISGRI all-sky map in the 18-60 keV energy band, with (merged) sources detected in this work labelled.}
\label{fig:GC}
\end{figure*}
22 of the sources in cat1000 were labelled as blended, in other words their positions were considered unreliable due to nearby sources within the angular resolution of the telescope. In some cases there was ambiguity in the identification of what appeared as one source, whereas other pairs of sources showed clear extension beyond a single point source. All cases were subject to lengthy visual inspection to determine their best representation in the final catalog. In addition to simple blended pairs, 13 sources in the crowded Galactic Centre region were also listed as blended - in this case the true determination of the emitters was impossible due to the crowded and highly variable nature of the region.
Some of these regions may prove useful test cases to understand if the Bayesian matching method is better able to untangle these complex regions, but unfortunately all of the blended sources in cat1000 are faint, with persistent fluxes below 10 mCrab. Nevertheless, some pairs are of sources that are variable in nature, so de-blending may be possible as precise positions with small uncertainties can be obtained from science windows during which the sources are bright. As our method searches for sources at the ScW-level, we frequently recover sources that may appear blended in a stacked image (see Figure~\ref{fig:GC}). If the point source location is sufficiently robust, these sources should not be merged during subsequent analysis.


\section{Conclusion and prospects for generating future catalogs}
\label{sec:conclusions}

Our CNN-based approach to source detection in IBIS/ISGRI has yielded several key advantages. 
Firstly, our method utilises all five energy bands simultaneously -- 
not only does this improve the accuracy and allow the network to detect sources at a lower flux threshold, it also speeds up the merging process as it results in a single excess for each astrophysical source instead of one from each energy band.
Secondly, the speed on which the CNN can be applied to the entire data set allows us now to look on a ScW timescale and detect sources that only appear in a single ScW but fall below the detection threshold in all-sky maps stacked from images spanning a
revolution time-scale.
Finally, our CNN-driven approach removes human biases from the source detection process, making our list of sources more impartial than previous approaches.

Bayesian matching recovers marginally more sources than {\em megamerge} - but has the `right' set of answers and has removed a bias from the process by not using the reference catalog as a starting point. Another limitation removed is that the order in which excesses are presented to the algorithm no longer impacts the end result.

Looking on a ScW level allows us to detect sources that have outbursts on smaller timescales than previous studies of the IBIS/ISGRI dataset. This approach also helps us to resolve the emission from the GC - detection at ScW level is easier to do than with stacked images as sources are not all 'on' at the same time. 

The combination of the CNN and Bayesian matching produces a very accurate merged list of detections with very few detections needing to be manually checked - compared to the old method which took 2.5 years for 9 people to manually check each source for inclusion into the catalog. 

If we wanted apply these tools to generate future catalogs we would also need to extend this work to include revolution maps and all-sky maps, enabling us to find the weak persistent sources that are not detected at a high enough significance in a single ScW. When the ScW images are stacked currently using a mosaic tool, the point spread functions (PSF) of the sources become distorted and we may find the CNN's performance drops when applied to these images. One way to overcome this would be to use HEALPix, which produces a subdivision of a spherical surface in which each pixel covers the same surface area as every other pixel. This would maintain the PS of the sources and allows us to use our CNN in order to source detect in stacked images although we would need to take care to maintain the same resolution. 

One possible setback of this method is that because the CNN looks on a local 11x11 pixel window it does not know how noisy the entire map is. In most cases this should not be a shortcoming, as the local noise level will be more important for informing the CNN of the likelihood of a source being present. However there are some isolated cases where the global noise map is useful. When domain experts make a decision about a source the 'flatness' of the map is taken into account. In addition to this, two false positives were still detected in the extremities of the partially coded FOV, although such excesses would normally be suppressed by the  low exposure in that area.  This problem could be overcome by not applying the CNN in the border region of the ScW maps, or a flag applied to any detections in this area for a manual inspect.  

Our newly developed source detection and merging method is reliable, scalable, removes need for continuous human intervention and eliminates some of the human subjectively that previously existed. This will be ideal for application to help generate future ISGRI catalogs.

\section*{Acknowledgements}
The authors would like to thank all the members past and present of the INTEGRAL IBIS survey team who have contributed to the catalogs produced so far. We thank them for valuable discussions around the methods used to date and how we might contribute to streamline this process in the future. We gratefully acknowledge the support of NVIDIA Corporation with the donation of the Titan Xp GPU used for this research.





\section*{Data Availability Statement}

The data underlying this article were accessed from the INTEGRAL Science Data Centre (www.isdc.unige.ch/integral), the derived survey data was published as 'The IBIS soft gamma-ray sky after 1000 INTEGRAL otbits' (DOI - 10.3847/0067-0049/223/1/15) and new models and datasets can be obtained by contacting the author.

\bibliographystyle{mnras}
\bibliography{faint_source_detection}

\bsp	
\label{lastpage}
\end{document}